\documentclass[twocolumn, superscriptaddress, showpacs, pra]{revtex4-1}

\global\arraycolsep=1pt
\usepackage{color,amsbsy,amssymb,latexsym,amsfonts, amsmath,cancel}
\usepackage{braket}
\usepackage{mathrsfs}
\usepackage{graphicx}
\usepackage{fancyhdr}

\newcommand{\bel}[1]{\begin{equation}\label{#1}}                     
\newcommand{\bal}[1]{\begin{eqnarray}\label{#1}}   
\newcommand{\be}{\begin{equation}}               
\newcommand{\ba}{\begin{eqnarray}}           
\newcommand{\ee}{\end{equation}}
\newcommand{\ea}{\end{eqnarray}}

\newcommand{\im}{\mathrm{i}}

\newcommand{\abs}[1]{\left| #1 \right|}
\newcommand{\pardif}[2]{\frac{\partial #1}{\partial #2}}
\newcommand{\bea}{\begin{equation}}
\newcommand{\eea}{\end{equation}}

\newcommand{\CReal}{\mathcal{C}}
\newcommand{\sgn}{\mathrm{sgn}}

\begin{document}


\title{Quantum holonomy in Lieb-Liniger model}

\author{Nobuhiro Yonezawa}
\email[Email:]{yonezawa@sci.osaka-cu.ac.jp}
\affiliation{Osaka City University Advanced Mathematical Institute (OCAMI), Sumiyoshi-ku, Osaka 558-8585, Japan}

\author{Atushi Tanaka}
\email[Email:]{tanaka-atushi@tmu.ac.jp}
\affiliation{Department of Physics, Tokyo Metropolitan University, Hachioji, Tokyo 192-0397, Japan}

\author{Taksu Cheon}
\email[Email:]{taksu.cheon@kochi-tech.ac.jp}
\affiliation{Laboratory of Physics, Kochi University of Technology, Tosa Yamada, Kochi 782-8502, Japan}

\begin{abstract}
We examine a parametric cycle in the $N$-body Lieb-Liniger model that starts from 
the
free system and goes through Tonks-Girardeau and super-Tonks-Girardeau regimes and comes back to the free system.  
We show the existence of  exotic quantum holonomy, whose detailed workings 
are
analyzed with the specific sample of two- and three-body systems.
The classification of eigenstates based on clustering structure naturally emerges from the analysis.   
\end{abstract}

\maketitle
\thispagestyle{fancy}

\section{Introduction}

Among the solvable models of quantum mechanics, the Lieb-Liniger system \cite{LL63} belongs to the selective class of models that are genuinely many-body.  It is a system made up of identical bosons interacting through two-body contact force.  It was later shown that the one-dimensional system of identical fermions with two-body contact interactions can be rigorously mapped to the Lieb-Liniger system with strong and weak coupling regimes interchanged \cite{Gi60, CS99}.
Several further extensions of the model with anyon statistics has been found \cite{Gu87,Ku99,OB06,BG08}, and they are also known to be mathematically equivalent to the original model.  The thermodynamics of the Lieb-Liniger model has been studied extensively \cite{YY69,Ya70,Wa02,IT12}.

What has made the Lieb-Liniger model a focus of renewed recent attention is its experimental realization in the form of Tonks-Girardeau gas \cite{Ol98, KW04,Pa04}.   It has been shown that the coupling strength of the Lieb-Liniger system can be experimentally controlled through the Feshbach resonance mechanism \cite{AB05}.  In recent experiments by Haller and collaborators \cite{Ha09,Ha10}, a smooth change of the coupling strength from large negative values to large positive values, where one finds the super-Tonks-Girardeau system \cite{BB05}, has been realized.

The continuous transition from a strongly repulsive to strongly attractive regimes of Lieb-Liniger model inspires us to propose following parametric cycle $\CReal$. 
We start with 
the
noninteracting limit, increase the coupling strength adiabatically, reaching the strongly attractive regime crossing the $\pm \infty$ coupling limit, then decrease the absolute value of negative coupling strength until it reaches the noninteracting limit again.  In this paper, we show that 
the
initial energy eigenstates of the cycle are different from the final eigenstates, although the initial and the final Hamiltonians are identical.

This phenomenon, the so-called {\it exotic} quantum holonomy, in which quantum eigenvalues and eigenstates do not come back to the original ones after a cyclic parameter variation \cite{CT09}, belongs to a wider class of quantum holonomy that comprises both the celebrated Berry phase \cite{Be84} and the Wilczek-Zee holonomy \cite{WZ84} which appears in systems with degenerate eigenvalues.
The exotic quantum holonomy in the $\delta$-function potential system was considered in \cite{Ch98}.  Here we report a finding of the quantum holonomy in 
{\it many-body systems} 
interacting through the $\delta$-function potential.

The plan of this paper is as follows.
In Sec.~\ref{sec:ForwardCycle}, we derive the spectral equation for Lieb-Liniger model in two different forms to demonstrate the presence of quantum holonomies with respect to $\CReal$.
In Sec.~\ref{sec:BackwardCycle}, we show that the backward cycle is not always possible due to the clustering of particles. This leads to the concept of minimal states, which we utilize to classify the spectrum of the system in Sec.~\ref{sec:example}.
In Sec.~\ref{sec:EP}, we provide another view of the quantum anholonomy by focusing on the two-body system through the complexification the coupling strength.
Section~\ref{sec:summary}  contains our conclusion.

\section{Adiabatic cycle $\CReal$ for Lieb-Liniger model}
\label{sec:ForwardCycle}

Let us consider $N$ bosons confined in 
a
one-dimensional
space. The system is described by the Hamiltonian
\begin{align}
\begin{split}
H
=-\frac{1}{2}\sum_{j=1}^N\pardif{^2}{x_j^2}
    +g\sum_{j=1}^N \sum_{l=1}^{j-1}\delta(x_j-x_l),
\end{split}\label{eq:LL_H}
\end{align}
where the unit is chosen such that $\hbar$ and the mass of a particle can be set to unity.  The parameter $g$ is the interaction strength.
We impose the periodic boundary condition 
to the position space.
For simplicity, $L=2\pi$ is assumed, 
where $L$ is the period in the position space.
It is straightforward to extend our analysis to an arbitrary $L$,
as long as $0 < L < \infty$.

We look at the dependence of eigenenergies and eigenvectors on the coupling strength $g$.  In
particular, we focus on the cycle $\CReal$, which 
consists of three stages $\CReal^{(s)}$ ($s=1,2,3$). 
In the first stage $\CReal^{(1)}$, 
$g$ is prepared to be $0$ and is adiabatically increased to $\infty$.
Next, in stage $\CReal^{(2)}$,
$g$ is suddenly flipped from $\infty$ to $-\infty$.
In the final stage $\CReal^{(3)}$, 
$g$ is again adiabatically
increased to $0$, which is the initial value of $g$.
We denote the initial and final points of $\CReal$ as $g=0$ and
$g=0-$, respectively, to distinguish them.

The eigenvalue problem of $H$, Eq.~\eqref{eq:LL_H}, can be solved by
the Bethe ansatz,
where an eigenfunction is composed of $N$ plane waves specified by
a set of quasimomenta, also called rapidity $k_j$, 
which
satisfy
\begin{align}
\begin{split}
&\exp \left(
       \im 2 \pi k_j 
    \right)
=\prod_{l\neq j}\frac{k_{jl}+\im g}{k_{jl}-\im g},    
\end{split}\label{eq:pre_spectrum_condition}
\end{align}
where 
$k_{jl}=k_j-k_l$~\cite{LL63}.
We examine how 
$k_j(g)$'s, 
which are chosen to be smooth as $g$ is
varied, are changed by the cycle $\CReal$. 
The function $k_j(g)$ completely
characterizes the parametric evolution of eigenenergies, as well as
the ``adiabatic'' evolution of eigenvectors along $\CReal$. 
The analysis is decomposed into the three
stages $\CReal^{(s)}$ ($s=1,2,3$).

At the initial point $g=0$ of the first stage $\CReal^{(1)}$,
$k_j(0)$ takes an integer value.
Without loss of generality, we can choose the order of 
$k_j(g)$'s
so as to satisfy 
$k_1(g) < k_2(g) < \dots < k_N(g)$
for small positive $g$~\cite{Do93}. This ensures 
$k_1(0) \le k_2(0) \le \dots \le k_N(0)$. 

We introduce two quantized quantities which is conserved during the parametric evolution of $k_j(g)$ along $\CReal^{(1)}$.
Such ``topological invariants'' provide a way to evaluate
the change of $k_j(g)$ induced by stage $\CReal^{(1)}$.

First, during the interval $0\le g < \infty$, we have an integer
\begin{align}
  \label{eq:defI}
  I_j(g)\equiv 
  k_j(g) - \frac{1}{\pi}\sum_{j\ne l}\arctan\frac{g}{k_{jl}(g)}.
\end{align}
This 
is a consequence of 
Eq.~\eqref{eq:pre_spectrum_condition} and 
$
({t+\im})/({t-\im})
=-e^{-2i\arctan t}
,
$
which is applicable as long as $t^{-1}\ne 0$.
We use the principal branch of $\arctan$ throughout this paper.
This is justified for Eq.~\eqref{eq:defI} because 
$g/k_{jl}(g)$
does not cross its standard branch cuts,
which emanate from $\pm i$ to $\pm i\infty$~\cite{atan}.
The continuity and discreteness of $I_j(g)$ in $0\le g < \infty$
imply that $I_j(g)$ takes a constant value,
which can be determined from the value of $k_j(g)$ at 
the initial point of $\CReal$, i.e., 
\begin{align}
I_j(g) = k_j(0)
.
\label{eq:continu_at_0}
\end{align}
Second, Eq.~\eqref{eq:pre_spectrum_condition} and another formula for
$\arctan$
$
  ({t+\im })/({t-\im})=e^{2\im\arctan\left(t^{-1}\right)}
,
$
which holds for $t\ne 0$, implies that,
in the interval $0 < g \le \infty$,
\begin{align}
  \label{eq:defJ}
  J_j(g)
  \equiv k_j(g) + \frac{1}{\pi}\sum_{l\ne j}\arctan\frac{k_{jl}(g)}{g}
\end{align}
is a half-integer for even $N$ and an integer for odd $N$~\cite{LL63}. 
Following a similar argument for $I_j(g)$ above,
we obtain the value of the invariant $J_j(g)$ for $0 < g \le \infty$:
\begin{align}
  J_j(g) = k_j(\infty)
  .
\end{align}

Now we evaluate the change of $k_j(g)$
during $\CReal^{(1)}$ using these invariants.
From Eqs.~\eqref{eq:defI} and~\eqref{eq:defJ}, we obtain
\begin{align}
  \label{eq:Delta1k}
  k_j(\infty) - k_j(0)
  = \frac{1}{2}\sum_{l\ne j}
  \sgn\Re\frac{k_{jl}(g)}{g}
  ,
\end{align}
where we used the identity
\begin{align}
\begin{split}
\arctan(t)+\arctan(1/t)=\frac{\pi}{2} \mathrm{sgn}[\Re(t)].
\end{split}\label{eq:form_arctan}
\end{align}
We note that the right-hand side of
Eq.~\eqref{eq:Delta1k} makes sense only for $0 < g < \infty$.
Here, $k_{jl}(g)$ is positive for $j>l$  and negative for $j<l$, since we have 
assumed the order of $k_j(g)$ at the initial point of $\CReal^{(1)}$, and
the sign of $k_{jl}(g)$ 
does not change for $g > 0$~\cite{Do93}.
This implies
\begin{align}
\sum_{l\ne j}
  \sgn\Re\frac{k_{jl}(g)}{g}
=\sum_{l=1}^{j-1}-\sum_{l=j+1}^{N}
.
\end{align}
Accordingly, we obtain
\begin{align}
  \label{eq:Delta1kResult}
  k_j(\infty) - k_j(0)
  = j - \frac{N+1}{2}.
\end{align}

Next we examine the second stage $\CReal^{(2)}$, where
$g$ suddenly changes from $\infty$ to $-\infty$.
Note 
that all $k_j(\infty)$'s are
finite because of Eq.~\eqref{eq:Delta1kResult}.
Since a finite root of the Bethe equation, Eq.~\eqref{eq:pre_spectrum_condition},
at $g=\infty$ is also its root
at $g=-\infty$,
we employ a smooth extension of $k_j(g)$ along $\CReal^{(2)}$, i.e.,
\begin{align}
\begin{split}
  \label{eq:kInfinityContinuity}
  k_j(-\infty) = k_j(\infty).
\end{split}
\end{align}
Details of the justification of our choice are explained in Appendix~\ref{app:prf_real_num}.

We further extend $k_j(g)$'s for the final stage $\CReal^{(3)}$.
First, we impose that $k_j(g)$'s satisfy $J_j(g) = k_j(\infty)$
within the interval $-\infty \le g < 0$. 
This implies that $k_j(g)$'s also satisfy 
Eq.~\eqref{eq:pre_spectrum_condition}.
We provide an argument that such $k_j(g)$'s exist for 
$-\infty \le g < 0$,
and are real-valued in Appendix~\ref{app:prf_real_num}
.
We accordingly conclude that $J_j(g)$ is independent of $g$ within the 
interval $-\infty \le g < 0$, because $k_j(g)$'s take real and finite 
values there.

Second, we examine $I_j(g)$ [Eq.~\eqref{eq:defI}] for 
$-\infty < g \le 0$. In contrast to the analysis of $J_j(g)$ above,
we need to inspect $k_j(0-)$, which is the final value of 
$k_j(g)$ in $\CReal^{(3)}$ and is different from 
the initial value $k_j(0)$. We carry this out by extending
$k_j(g)$'s from the interval $-\infty \le g < 0$.
We explain 
the details
of our argument in Appendix~\ref{app:cont_at_0-} and
only show the result that $I_j(g)$ agrees with $k_j(0-)$
within the interval $-\infty < g \le 0$.

%
%
%
%
The change of $k_j(g)$ in the path $\CReal^{(3)}$ is given by
\begin{align}
  k_j(0-)- k_j(-\infty)
  = -\frac{1}{2}\sum_{l\ne j}
  \sgn \Re
  \frac{k_{jl}(g)}{g}
  .
\end{align}
We can ensure that
\begin{align}
k_1(g)<k_2(g)<\cdots<k_N(g)
\end{align}
because it holds at $g=-\infty$
(see 
Appendix 
\ref{app:cont_at_0-}).
Recalling the fact that $g$ is negative here, we obtain
\begin{align}
  \label{eq:Delta3kResult}
  k_j(0-)- k_j(-\infty)
  = j - \frac{N+1}{2}
  .
\end{align}

Combining above three arguments, we obtain 
a nontrivial
change of $k_j(g)$ due to $\CReal$ in the form
\begin{align}
  k_j(0-)- k_j(0)
  = 2j - (N+1)
  .
\end{align}
Note that
the total momentum remains unchanged during the cycle $\CReal$.
The final energy and state after the adiabatic 
cycle, 
however, are
different from the initial ones, showing that $\CReal$ 
induces the eigenenergy and eigenspace anholonomies~\cite{Ch98}.
We also remark that $k_1<k_2<\dots <k_N$ holds at the end of $\CReal$.
This implies that we can repeat the adiabatic cycle $\CReal$ 
arbitrarily, and the repetition of $\CReal$ will induce
the further instances of the eigenenergy and eigenspace anholonomies.

We can summarize our results in terms of a mapping between
two sets of quasimomenta of free bosons, i.e., 
$k_j(0)$'s and $k_j(0-)$'s .
It is sufficient to consider the case that initial condition  $n_j\equiv k_j(0)$ satisfies
$n_1 \le n_2 \le \dots \le n_N$.
With the notation $n_j'\equiv k_j(0-)$, the mapping 
 $(n_1, n_2, \dots, n_N)
 \mapsto (n_1', n_2', \dots, n_N')=F(n_1, n_2, \dots, n_N)$, which is given by
\begin{align}
  \label{eq:F}
  &
F(n_1, n_2, \dots, n_N)
     \nonumber \\ 
  &
  = (n_1 -N+1, n_2-N+3, \dots, n_N+N-1)
  ,
\end{align}
expresses the quantum holonomy 
induced by the cycle $\CReal$.

\section{Inverse cycle}
\label{sec:BackwardCycle}

We now examine
the inverse of the cycle $\CReal$.
In contrast to the forward cycle $\CReal$, the parametric variation along
the inverse ${\CReal}^{-1}$
is not always possible.
This is because the clustering of particles at $g=-\infty$ induces
the divergence of eigenenergy~\cite{LL63}.
Such a clustering invalidates the use of the Hamiltonian, Eq.~\eqref{eq:LL_H}.
We call an eigenstate of free boson at $g=0$ a {\em minimal state}
if the the parametric variation along ${\CReal}^{-1}$ is impossible.
The precise condition for appearance of the minimal state is the subject of this section.

Formally, ${\CReal}^{-1}$ corresponds to the inverse of the mapping 
$F$ [Eq.~\eqref{eq:F}] on the sets of quasimomenta at $g=0$:
\begin{align}
  \label{eq:InvF}
  &
  F^{-1}(n_1, n_2, \dots, n_N)
  \nonumber \\ 
  &
  = (n_1 +N-1, n_2+N-3, \dots, n_N-N+1)
  ,
\end{align}
where we impose the ordering condition $n_1\le n_2 \le \dots \le n_N$.
When the distance between $n_j$'s are far enough,
$F^{-1}$ preserves the ordering.
This is the case that ${\CReal}^{-1}$ can be realized, and
the resultant energy and quantum state are the solution of the eigenvalue
problem of $H$ [Eq.~\eqref{eq:LL_H}] at $g=0$.
On the other hand, 
when a pair of $n_j$'s 
is too close,
$F^{-1}$ breaks the ordering, which implies the emergence of 
the clustering of particles during the inverse cycle.
There are two possible cases.
The first  case is where a pair of quasimomenta,
say, $n_j$ and $n_{j+1}$,
are degenerate, i.e., $n_j = n_{j+1}$. 
By applying $F^{-1}$,
the resultant quasimomenta satisfy $n_j > n_{j+1}$.
In fact, the eigenenergy diverges $-\infty$ as $g\to-\infty$
during ${\CReal}^{-1}$.
The second case, $n_j = n_{j+1}+1$, also leads 
the clustering of particles.

The argument above is sufficient to determine the condition for the minimal states. 
When 
there is, at least, a pair of two quasimomenta at $g=0$
that satisfies 
\begin{align}
  \label{eq:conditionMinimalStates}
  |n_j-n_{j+1}| \le 1
  ,
\end{align}
states specified by $n_j$ and $n_{j+1}$ are minimal states.

\section{Classification of spectra}
\label{sec:example}

%
Because of the existence of quantum holonomy, some states are reachable by 
the repetitions of
parametric cycles $\CReal$ and ${\CReal}^{-1}$ starting from one particular eigenstate,  while other states are not.  This offers the classification of whole eigenstates into families of states connected by quantum  holonomy.  Such a family can be specified by a minimal state introduced above, because an arbitrary eigenstate with a finite energy can become minimal by a finite repetition of ${\CReal}^{-1}$.

From one minimal state, we can find other minimal states using the symmetries of the Hamiltonian~\eqref{eq:LL_H}.
Suppose that a minimal state is specified by quasimomenta $(n_1,n_2,\dots,n_N)$.
The translational symmetry implies that $(n_1+1,n_2+1,\dots,n_N+1)$ is also a minimal state, whose total momentum is larger by $N$ than the original one.
For an arbitrary integer $\ell$, $(n_1+\ell,n_2+\ell,\dots,n_N+\ell)$ is also a minimal state. 
The reflection symmetry implies that  $(-n_N, \dots, -n_2, -n_1)$ is also a minimal state, which may or may not be different from the original state.

\begin{figure}[tb]
  \begin{center}
    \includegraphics[keepaspectratio=true,width=0.5\textwidth]{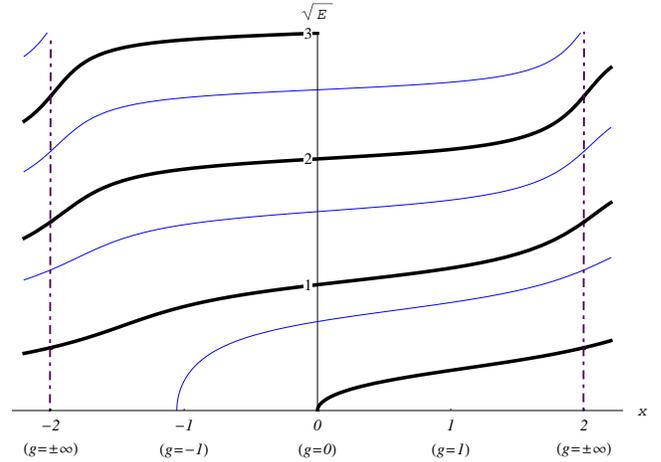}
  \end{center}
  \caption{
    (Color online)
    Parametric evolution of eigenenergies of the two-body Lieb-Liniger model,
    where the $x$ and $y$ axes indicate $(4/\pi)\arctan g$ and $\sqrt{E}$,
    respectively.
    The unit is chosen such that $\hbar$ and the mass of a particle are set to 
    unity. The period of the position space is chosen to be $2\pi$.
    The thick (black) and thin (blue) lines correspond to the families specified
    by the minimal states $(0,0)$ and $(0,1)$, respectively.
    See Eqs.~\eqref{eq:00} and \eqref{eq:01}.
    Note that the eigenenergies are continuous at $g=\pm\infty$.
  }
  \label{fig:N2}
\end{figure}

Hence, it is sufficient to find all minimal states whose total momenta
satisfy the condition
\begin{align}
-\frac{N}{2} < \sum_j n_j \le \frac{N}{2},
\end{align} 
to  enumerate all minimal states using the translational symmetry, offering a way to classify the spectra of the Lieb-Liniger model completely.  We illustrate this classification for few-body cases.

We start the analysis of $N=2$ case with two minimal states, 
\begin{align}
  \label{eq:minimal2}
  (0,0)\quad\text{and}\quad(0,1)
  .
\end{align}
We obtain two families of eigenstates at $g=0$ from 
these two minimal states, by repeating $\CReal$,
\begin{align}
  \label{eq:00}
  (0,0)\mapsto (-1,1)\mapsto(-2,2)\mapsto \cdots
  ,\intertext{and}
  \label{eq:01}
  (0,1)\mapsto (-1,2)\mapsto(-2,3)\mapsto \cdots
  ,
\end{align}
respectively. 
The eigenenergies of these families are depicted in Fig.~\ref{fig:N2}.  By shifting the total momentum from the two minimal states [Eq.~\eqref{eq:minimal2}], we obtain an infinite number of minimal states $(\ell,\ell)$ and $(\ell,\ell+1)$ with an arbitrary integer $\ell$.
The 
$(\ell,\ell)$- and $(\ell,\ell+1)$-families have the set of quasimomenta at $g=0$ given by $\{(\ell-m,\ell+m)\}_{m=0}^{\infty}$ and $\{(\ell-m,\ell+1+m)\}_{m=0}^{\infty}$, respectively.
This exhausts the minimal states and families for $N=2$.

\begin{figure}[tb]
  \begin{center}
    \includegraphics[keepaspectratio=true,width=0.5\textwidth]{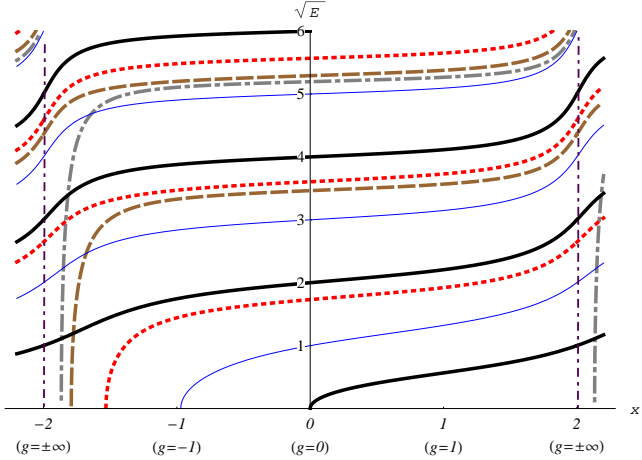}
  \end{center}
  \caption{
  (Color online)
  Eigenenergies of $N=3$ families, where the $x$ and $y$ axes are 
  the same as in Fig.~\ref{fig:N2}.
  The total momentum of all families shown here is zero. 
  The thick (black) line 
  corresponds to the family [Eq.~\eqref{eq:000}]
  specified by the minimal state $(0,0,0)$.
  The thin (blue) line corresponds to the $(-1,0,1)$ family.
  These two families are trimers.
  The dotted (red), dashed (brown), and dash-dotted (gray) lines are dimer
  families specified by minimal states $(-1,-1,2)$, $(-2,-2,4)$, and $(-3,-3,6)$,
  respectively.
  Although there are level crossings, the adiabatic theorem ensures
  that the adiabatic time evolution is confined within 
  a family~\cite{Kato-JPSJ-5-435}.
  The choice of the unit is
the
same as in Fig.~\ref{fig:N2}.
  }
  \label{fig:N3}
\end{figure}

\begin{figure}[tb]
  \begin{center}
    \includegraphics[keepaspectratio=true,width=0.5\textwidth]{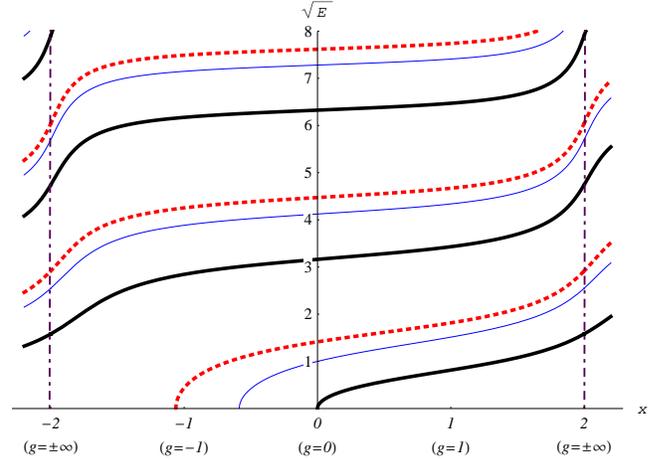}
  \end{center}
  \caption{
  (Color online)
  Parametric evolution of eigenenergies of the $N=4$ case, 
  where the $x$ and $y$ axes are 
  the same as in Fig.~\ref{fig:N2}.
  The thick (black) line 
  corresponds 
  to the family
  $(0,0,0,0)\mapsto(-3,-1,1,3)\mapsto(-6,-2,2,6)\dots.$
  The thin (blue) and dotted (red) lines correspond to 
  $(-1,0,0,1)$ and $(-1,-1,1,1)$ families, respectively.
  %
  The choice of the unit is 
the
same as in Fig.~\ref{fig:N2}.
  }
  \label{fig:fig_4body.eps}
\end{figure}

The $N=3$ case is far more complex than the $N=2$ case.
First, we consider the case that the total momentum is zero, where an infinite number of minimal states can be found. We
depict some of them in Fig.~\ref{fig:N3}.
There are two minimal states,
\begin{align}
  (0,0,0)\quad\text{and}\quad(-1,0,1)
  ,
\end{align}
which are called 
trimers~\cite{MS98}, 
because the clustering of all three 
particles occurs in the limit $g\to-\infty$. 
The family of eigenstates at $g=0$ specified by the minimal
state $(0,0,0)$ is
\begin{align}
  \label{eq:000}
  (0,0,0)\mapsto (-2,0,2)\mapsto(-4,0,4)\mapsto\dots
  .
\end{align}
Besides, there are an infinite 
number of minimal states,
\begin{align}
  \{(-\ell,-\ell, 2\ell)\}_{\ell>0}
  \quad\text{and}\quad
  \{(-2\ell,\ell, \ell)\}_{\ell>0}
  ,
\end{align}
where the latter set can be induced through the use of the reflection 
symmetry. These minimal states are called 
dimers~\cite{MS98}, 
because the 
clustering of two particles occurs in the limit $g\to-\infty$. 
Second, we consider the case $\sum_j n_j = 1$. We have a trimer,
\begin{align}
(0,0,1) ,
\end{align}
and an infinite number of dimers
\begin{align}
&
\{(-\ell,-\ell, 2\ell+1)\}_{\ell>0}, \quad
\{(-\ell,-\ell+1, 2\ell)\}_{\ell>0},
\nonumber \\
&
\{(-2\ell+1,\ell, \ell)\}_{\ell>0}, \quad
\{(-2\ell,\ell, \ell+1)\}_{\ell>0}.
\end{align}
Note that all minimal states that satisfy $\sum_j n_j = -1$ 
can be obtained from the minimal state with $\sum_j n_j = 1$ 
through the use of the reflection symmetry. 
We obtain all other minimal states from above using the translational and reflection symmetry.

It is possible to enumerate minimal states and associated spectral families
in a similar way for larger $N$. We simply close this section by showing several 
families of the $N=4$ system in Fig.~\ref{fig:fig_4body.eps}.

\section{Exceptional points}
\label{sec:EP}
So far we have focused on the quantum  holonomy induced by the real 
cycle $\CReal$.
In this section, we examine the relationship between the exotic quantum
holonomy and non-Hermitian degeneracy points, which are also known as
Kato's exceptional points~\cite{KatoExceptionalPoint,Heiss:CzecJP-54-1091},
using the complexification of the coupling parameter $g$.
When we adiabatically vary $g$ along a cycle that
encloses
an exceptional 
point, the permutation of eigenenergies as well as eigenspaces
occurs.
This resembles the exotic quantum holonomy.
Indeed, in 
Ref.~\cite{Ki10} it is argued that, through an analysis of a 
quantum kicked top,
the quantum holonomy has a correspondence
with the exceptional points.
In other words, it is conjectured that 
the eigenenergy and eigenspace anholonomy can be understood as 
a result of the metamorphosis of eigenenergies and eigenstates
induced by the encirclements around the exceptional points.
In the following, we offer another example of this conjecture using 
the two-body Lieb-Liniger model by deforming $\CReal$ in
the complexified $g$ space.

\begin{figure}[h]
  \centering

  \begin{minipage}[h]{0.23\textwidth}
    \includegraphics[width=\textwidth]{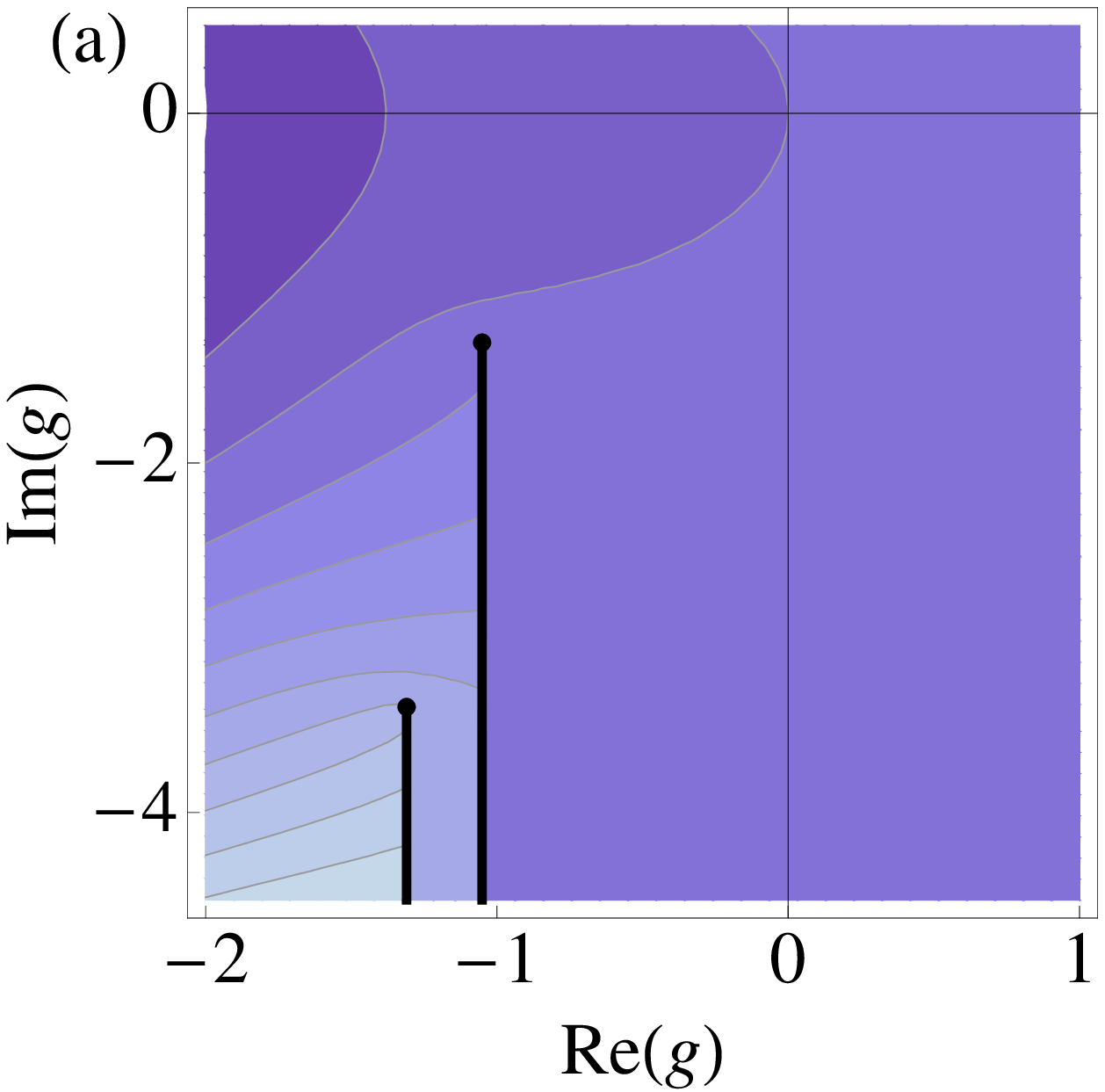}    
  \end{minipage}
  \
  \begin{minipage}[h]{0.23\textwidth}
    \includegraphics[width=\textwidth]{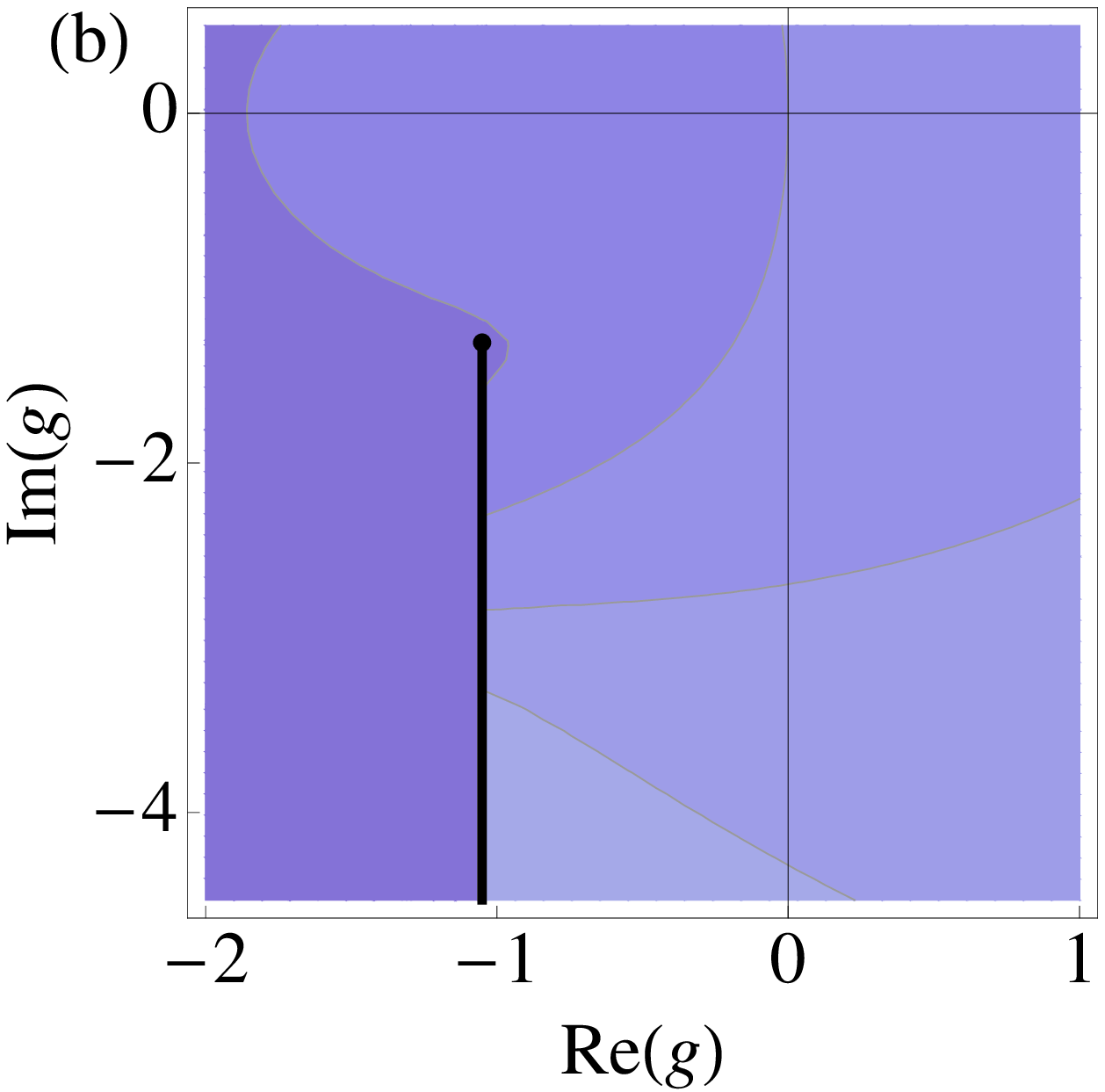}    
  \end{minipage}
  \\[\baselineskip]

  \begin{minipage}[h]{0.23\textwidth}
    \includegraphics[width=\textwidth]{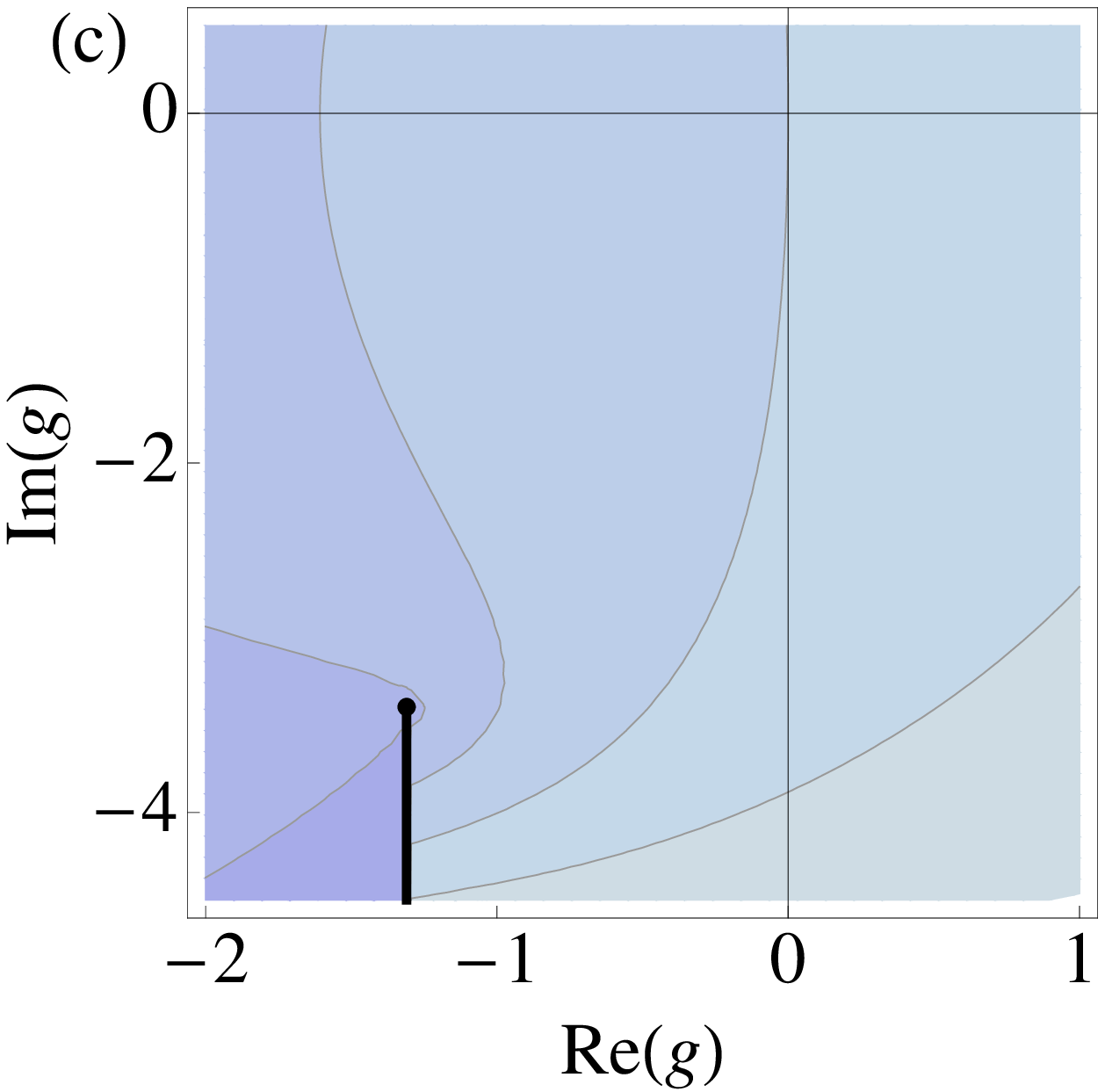}    
  \end{minipage}
  \ \ 
  \begin{minipage}[h]{0.23\textwidth}
      \includegraphics[width=\textwidth]{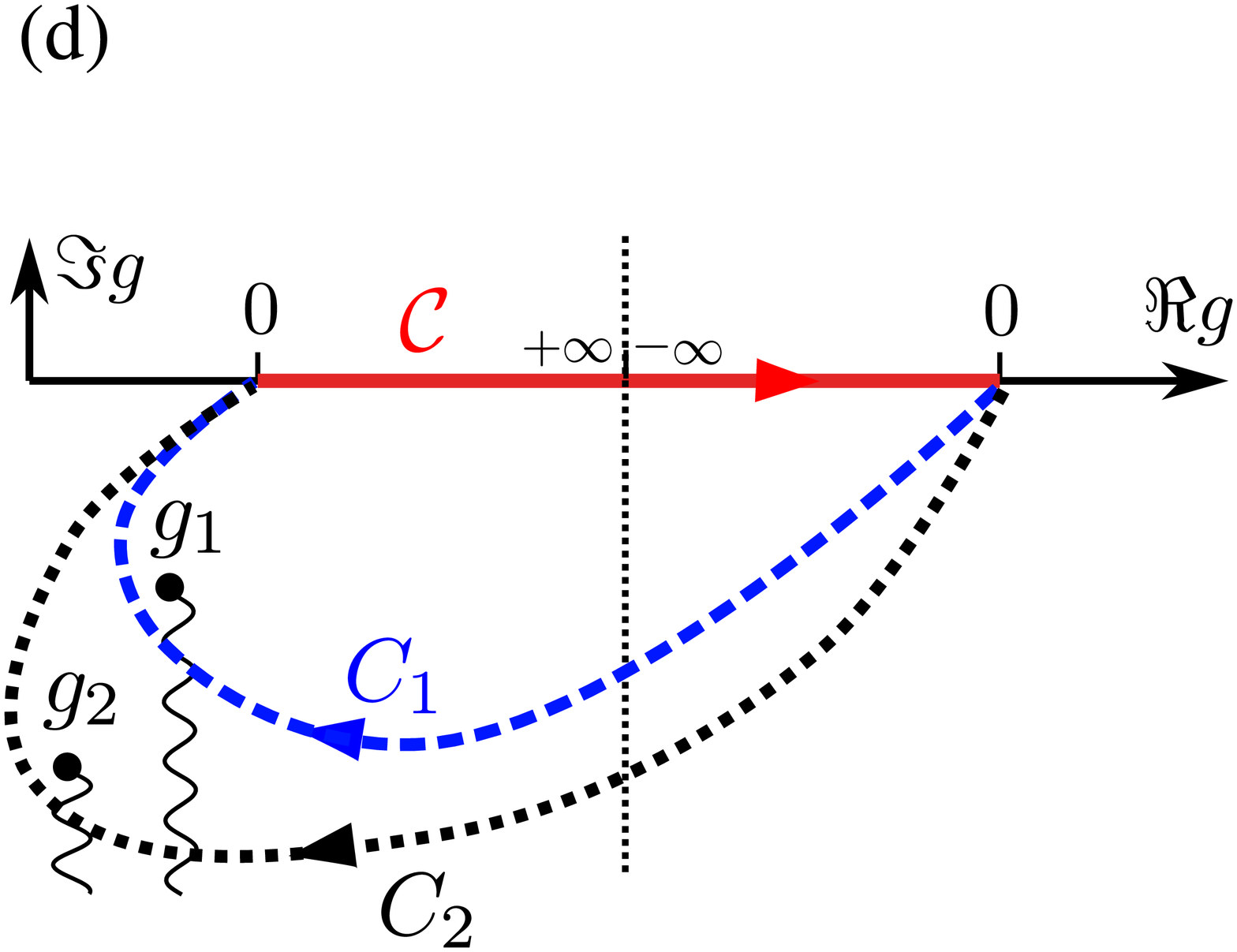}    
  \end{minipage}

  \caption{%
    (Color online)
    Contour plots of $\Re E_n(c)$:
    (a) $n=0$; (b) $n=1$; (c) $n=2$.
    Lighter (darker) color indicates larger (smaller) value of $\Re E$.
    Thin lines are the contours of $\Re E_n(c)$.
    The exceptional points are indicated by solid circles.
    Bold lines indicate the branch cuts.
    While all complex exceptional points appear in $E_0(c)$,
    each $E_n(c)$ ($n=1,2$) has a single exceptional point.
    (d) Schematic explanation of complex cycles that
    enclose exceptional points. 
    We depict $\CReal$ by a thick (red) line.
    Dashed (blue) and dotted (black) curves indicate
    $C_1$ and $C_2$, respectively. 
    See the main text.
    The choice of the unit is
the
same as in Fig.~\ref{fig:N2}.
  }
  \label{fig:ReEsheets}
\end{figure}

Due to the complexification of $g$, the Lieb-Liniger 
Hamiltonian~\eqref{eq:LL_H}
becomes
non-Hermitian, which
describes a one-dimensional dissipative Bose system~\cite{Du09}.
We obtain eigenenergies with complex-valued
coupling parameter $g$ through numerical computation.
We here focus on the $(0,0)$ family [Eq.~\eqref{eq:00}].
Let $E_n(g)$ denote the eigenenergy of the state whose quasimomenta
take $(-n,n)$ at $g=0$.
We depict $E_n(g)$ for $n=0,1,2$
in Fig~\ref{fig:ReEsheets}. We find that these eigenenergies compose a
Riemann surface. Its Riemann sheets $E_n(g)$ are connected by
the exceptional points and associated branch cuts 
(cf. Ref.~\cite{Heiss:JMP-32-3003}).

Under
the present choice of the branch cuts, all exceptional points
of the $(0,0)$ family appear in the $E_0(g)$ sheet. A pair of 
eigenenergies
$E_n(g)$ ($n>0$) and $E_0(g)$ has a pair of degenerate points
$g_n$ and $g_n^*$, where we choose $\Im g_n < 0$. 
We find that all degenerate points are of degree two.
Hence, the pair of eigenenergies for an exceptional point 
exhibits square-root-type singularity.
The encirclement around the exceptional 
point $g_n$ in the complex $g$ plane induces the permutation of 
$E_0(g)$ and $E_n(g)$.
We numerically confirm these properties of $g_n$ with
$n=1,2, \dots, 10$. 
We find that $\Re g_n$ and $\Im g_n$ 
decrease monotonically as $n$ increases.
We also obtain 
similar results
for the $(0,1)$ family.

We note that 
our numerical finding can be explained by a perturbation expansion 
for $g=-\infty$ with a small parameter $g^{-1}$,
as for the exceptional points that are far from 
the real axis~\cite{Ushveridze-JPA-21-955}.
We will explain the details in a forthcoming publication~\cite{TYC130}.

Let us consider the cycle that is a concatenation of 
$\CReal$ and $C_1$ in Fig~\ref{fig:ReEsheets} (d).
Because this cycle encircles the exceptional point 
$g_1$, 
the cyclic permutation $(E_0, E_1)$ occurs.
On the other hand, the cycle composed of $\CReal$ and $C_2$
induces 
the cyclic permutation among $(E_0, E_1, E_2)$.
As the cycle involves more deeper exceptional points,
the accuracy of the the resultant permutation become better 
to approximate a shift to eigenenergies
$(E_0, E_1, \dots)\mapsto (E_1, E_2, \dots)$,
which is realized by the quantum holonomy along the cycle $\CReal$.
In this sense, we may say that the spectrum of Lieb-Liniger model
feels the exceptional points that reside in the complex parameter
space to induce the quantum holonomy along $\CReal$.

\section{Conclusion}
\label{sec:summary}

We have shown in this work that an eigenstate of the free Lieb-Liniger  system $g=+0$ is transformed to another eigenstate with higher energy in the process of eigenspace anholonomy involving the parametric cycle $g: +0 \to +\infty:-\infty \to  -0$.  Experimental testing should be within the range of current techniques \cite{Ha09,Ha10}.  On the way to prove the existence of quantum holonomy, we have demonstrated that the eigenstates of the Lieb-Liniger model can be classified according to their clustering property.  The two- and three-boson systems have been analyzed in detail.

Our result can be interpreted in terms of geometry. 
Consisting of real numbers and $\pm\infty$, 
the
parameter space of coupling 
strength
is homeomorphic to $S^1$. 
Therefore, our anholonomy is affected by the topology of $S^1$. 
The presence of two kinds of invariants, $I_j(g)$ [Eq.~\eqref{eq:defI}]
and $J_j(g)$ [Eq.~\eqref{eq:defJ}], for the parametric evolution of
$k_j(g)$
reflects
the fact that at least two charts are required for $S^1$. 
Converting one of spectrum condition to the other by using the formula of arctan \eqref{eq:form_arctan} corresponds to coordinate transformation.
The cycle of the winding number $m$,
${\CReal}^m$,  increases $k_j(0)$ by
$m[2j - (N+1)]$.

The topological nature of the the quantum holonomy implies that it is stable against, at least, small perturbations~\cite{Tanaka-PRL-98-160407}. This also suggests that an experimental realization of the quantum anholonomy is possible in one-dimensional bosonic systems.

\section*{Acknowledgement}

This research was supported  by the Japan Ministry of Education, Culture, Sports, Science and Technology under the Grant numbers 22540396 and 24540412.

\appendix

\section{%
  Extension of $k_j(g)$'s to $-\infty \le g < 0$
}
\label{app:prf_real_num}

We examine $k_j(g)$'s that 
satisfy
\begin{align}
k_j(\infty)=k_j(g)+\frac{1}{\pi}\sum_{l\neq j}\arctan\frac{k_{jl}(g)}{g}
\label{eq:be_infty}
\end{align}
in the interval $-\infty \le g < 0$ in this appendix. 
Our argument consists
of two parts. First, we provide an argument that 
$k_j(g)$'s are real and finite for $-\infty \le g < 0$.
Second, we explain that such $k_j(g)$'s are the 
smooth extension of the ones defined in the first stage $\CReal^{(1)}$. 

We have already examined $k_j(\infty)$, which appears in the left-hand side of
Eq.~\eqref{eq:be_infty}, in the main text. In particular,
$k_j(\infty)$'s are real and finite. Also, $k_j(\infty)$'s are not 
degenerate, i.e.,
\begin{align}
\begin{split}
k_j(\infty)-k_{l}(\infty)
>0,
\label{eq:dif_k_infty}
\end{split}
\end{align}
for $j>l$, which is ensured by Eq.~\eqref{eq:Delta1kResult}.

We introduce an assumption that plays the crucial role in the following 
argument. We assume that there uniquely 
exists
$\{k_j(g)\}_{j=1}^N$ that 
satisfies Eq.~\eqref{eq:be_infty}. We note that
this assumption indeed holds, as for $g>0$~\cite{YY69}.

We show that $k_j(g)$'s are real numbers by \textit{reductio ad absurdum}. 
Namely, we suppose that $k_j(g)$ is not real and satisfies 
Eq.~\eqref{eq:be_infty} for a given $j$.
Accordingly, 
%
its complex conjugate $k_j(g)^*$ 
also satisfies Eq.~\eqref{eq:be_infty},
because Eq.~\eqref{eq:be_infty} is invariant under the complex conjugate.
Since 
$k_j(g)^*$ is different from $k_j(g)$ and there uniquely exists
$\{k_j(g)\}_{j=1}^N$,
there exists $j'$ such that $k_{j'}(g)=k_j(g)^*$ and $j'\ne j$.
We compare $k_j(\infty)$ and $\{k_{j'}(\infty)\}^*$, which are real numbers.
Using Eq.~\eqref{eq:be_infty}, we find $k_j(\infty)-\{k_{j'}(\infty)\}^* = 0$, 
which contradicts Eq.~\eqref{eq:dif_k_infty}.

A corollary of the above proposition, i.e., $k_j(g)$ are real for 
$-\infty\le g <0$, is the continuity of $k_j(g)$'s in the stages 
$\CReal^{(2)}$ as well as $\CReal^{(3)}$, as mentioned in the main text 
[see Eq.~\eqref{eq:kInfinityContinuity}]. In this sense, $k_j(g)$'s 
that satisfy Eq.~\eqref{eq:be_infty}
are
the smooth extension of
$k_j(g)$ for $0 < g \le\infty$.
We prove this corollary.
Since $k_j(g)$'s are real numbers,
we have
\begin{align}
\begin{split}
\abs{\arctan \left[k_{jl}(g)/g\right]} <\pi/2,
\end{split}
\end{align}
which implies 
that $k_j(g)$'s are finite, i.e.,
\begin{align}
\begin{split}
\abs{k_j(g)}&\le \abs{k_j(\infty)}+\frac{1}{\pi}\sum_{l\neq j}\frac{\pi}{2}<\infty,
\end{split}
\end{align}
where we use Eq. \eqref{eq:be_infty}.    
Hence, we find
\begin{align}
\lim_{g\to - \infty} k_{jl}(g)/g=0.
\end{align}
Taking the limit of Eq. \eqref{eq:be_infty} as $g\to-\infty$, we obtain
\begin{align}
\begin{split}
\lim_{g\to-\infty}k_j(g)
&=k_j(\infty)+\lim_{g\to - \infty}\frac{1}{\pi}\sum_{l\neq j}\arctan\frac{k_{jl}(g)}{g}\\
&=k_j(\infty).
\end{split}
\end{align}
Hence, we conclude that $k_j(g)$ is continuous at $g=-\infty$.
A similar argument above tells us that $k_j(g)$ is also
continuous at $g = \infty$.

\section{%
  Extension of $k_j(g)$'s from $g < 0$ to $g=0$
}
\label{app:cont_at_0-}
We have explained the smooth extension 
of
$k_j(g)$'s through the flip of 
$g$ from $\infty$ to $-\infty$ in
Appendix~\ref{app:prf_real_num}. Here we extend further $k_j(g)$'s from $g<0$
to $g=0$ to complete the analysis of the stage $\CReal^{(3)}$.  We
carry this out by showing $k_j(0-)= I_j(g)$.

%
%
%

To prepare this, we show that 
$k_j(g) \ne k_{j'}(g)$ holds for $g <0$, and
an arbitrary pair of $(j,j')$. We prove this by contradiction.
Suppose that there exists $g$ ($<0$), where
$k_j(g)=k_{j'}(g)$ ($j\ne j'$). 
Then
Eq.~\eqref{eq:be_infty} implies that 
$k_j(\infty)=k_{j'}(\infty)$,
which is inconsistent with Eq.~\eqref{eq:dif_k_infty}.

Next we show that $k_j(g) \ne k_{j'}(g)$ ($j \ne j')$ also holds 
in the limit $g\to 0-$.
We show this 
by using \textit{reductio ad absurdum}.
Suppose
$k_j(0-)= k_{j'}(0-)$.
We can assume $j>j'$ without loss of generality.
Hence, Eq.~\eqref{eq:be_infty} under the limit $g\to 0-$ implies
\begin{align}
\begin{split}
k_j(\infty)-k_{j'}(\infty)
=\lim_{g\to 0-}\frac{2}{\pi}\arctan \frac{k_{jj'}(g)}{g}
.
\end{split}
\end{align}
%
Thus we conclude $k_j(\infty)-k_{j'}(\infty) \le 0$, since
$\arctan\left[k_{jj'}(g)/g\right] < 0$ 
holds as long as $g<0$.
This conclusion contradicts
with Eq.~\eqref{eq:dif_k_infty} with $j>j'$.
We thus show $k_j(0-)\neq k_{j'}(0-)$.

We examine $I_j(g)$ [Eq.~\eqref{eq:defI}] in the limit $g\to 0-$.
Since $k_{jl}(0-)\ne 0$ holds, as shown above, we find
$\lim_{g\to 0-}~g/k_{jl}(g)=0$.
Hence, we obtain $I_j(0-)=k_j(0-)$. Because $I_j(g)$ 
is independent of $g$ for $-\infty < g < \infty$,
we conclude $I_j(g)=k_j(0-)$ for $g \le 0$.
We note that this result
and Eq.~\eqref{eq:continu_at_0} imply continuity of $k_j(g)$ at $g=0\pm$.
As for the proof of the continuity at $g=0+$, we refer 
to
Ref.~\cite{Do93}.

Finally, we remark that the present argument
and Eq.~\eqref{eq:dif_k_infty} imply
that the ordering of $k_j(g)$ satisfies
$k_1(g) < k_2(g) < \dots < k_N(g)$ for $g \le 0$.

\end{document}